# Critical Diameter for Continuous Evaporation is between 3 nm and 4 nm for Hydrophilic Nanopores

Sumith Yesudasan[*]

Department of Engineering Technology, Sam Houston State University, Huntsville, TX, USA 77341

*Corresponding author E-mail: sumith.yesudasan@shsu.edu

## Abstract

Evaporation studies of water using classical molecular dynamics simulations are largely limited due to their high computational expense. This study addresses that issue by developing coarse-grained molecular dynamics models based on Morse potential. Models are optimized based on multi-temperature and at room temperature using machine learning techniques like genetic algorithm, Nelder-Mead algorithm, and Strength Pareto Evolutionary Algorithm. The multi-temperature based model named as Morse-D is found to be more accurate than single temperature model in representing the water properties at higher temperatures. Using this Morse-D water model, evaporation from hydrophilic nanopores with pore diameter varying from $2\ nm$ to $5\ nm$ is studied. Our results show that the critical diameter to initiate continuous evaporation at nanopores lies between $3\ nm$ and $4\ nm$. A maximum heat flux of $21.3\ kW/cm^2$ is observed for a pore diameter of $4.5\ nm$ and a maximum mass flow rate of $16.2\ ng/s$ for a pore diameter of $5\ nm$. The observed heat flux is an order of magnitude times larger than the currently reported values from experiments in the literature for water, which indicates that we need to focus on nanoscale evaporation to enhance the critical heat flux.

## Introduction

The energy consumption has significantly increased over the past decade due to the increasing number of supercomputers and data center-based cloud computing. This trend will continue for decades which will put a strain in the energy sector. This trend also necessitates compact designs with higher energy density and hence calls for highly efficient and faster cooling systems. The cooling system of a data center typically consumes around 40% of the energy input[1]. This projection forecasts to approximately $150 billion per year cost by the year 2030 [2] and hence research efforts targeting the development of efficient cooling systems is vital in the fields of heat transfer and semiconductors. Among the conventional cooling techniques, evaporative water cooling emanates as a promising candidate due to its high specific heat capacity and high enthalpy of vaporization. Based on kinetic theory,[3] an evaporative heat flux of $20,000\ W/cm^2$ can be achieved using water at 1 atm. However, the experiments show that maximum values of CHF is largely limited in the range of $350{\sim}700\ W/cm^2$ despite of utilizing thin film evaporation[4,5]. This calls for improving our understanding of nanoscale and mesoscale characteristics of water evaporation. This paper is organized as follows. The first half of the paper will describe the necessity, development, and validation of the coarse-grained water models for evaporation studies. The second half will explain the application of such developed model to study the nanopore evaporation and its characteristics.

Past two decades have seen a plethora of research activities in utilizing nanoscale surface structures to enhance the heat transfer from a solid surface to liquid[6–10]. The studies ranged from theoretical modeling of nanopore evaporation using continuum methods[7,9] for $d = 10{\sim}100\ nm$, to experimental studies of nanopore evaporation[6,8] for $d = 24{\sim}100\ nm$, suggesting that high heat fluxes are possible through nano porous evaporation. But these enhancements are frequently limited by the critical heat flux (CHF) [11], contamination[7] or due to lack of continuous evaporation at nanopores and structures called dry out or burnout conditions. The evaporation characteristics of nanopores with $d < 10\ nm$ is not studied through simulations or experiments. While the experimental techniques pose challenges to validate and characterize evaporation at this length and time scale, computational tools such as molecular dynamics[12–14] can provide promising results without the need for expensive device fabrication and instrumentation. That being said, there is no water model that can capture or simulate all of its properties simultaneously [15]. Even the best performing and widely used models such as SPCE [16] and TIP3P [15] are unsuitable to study water evaporation, due to the excessive computational cost (refer supplementary material for calculation). This shifts our focus to computationally efficient models like coarse grain molecular dynamics (CGMD) models[17].

In a typical CGMD model, one or more water molecules are combined into a bead or a super-molecule to represent the bulk properties of the system. The field of coarse graining is a large research area and hence the readers are advised to refer to consolidated reviews found elsewhere in the literature [18–20]. Among various types of CGMD models, mono-bead models are appealing due to their low computational power consumption [21–23]. Specifically, the generalized Mie potential based water model [23], which maps one water molecule to a single bead, can accurately represent density, surface tension, and enthalpy of vaporization of water for a wide range of temperatures. But in their model the parameters are temperature dependent which makes them unsuitable for non-equilibrium molecular dynamics simulations. In the later studies, Lennard-Jones (LJ) potential which is a special case of Mie potential, is used to map four water molecules to a bead to form the classical MARTINI model[21,22] which is computationally efficient than the former. Most of these existing CGMD models are developed for bio-molecular studies [21,24,25] and have not been tested for heat transfer studies. Due to this reason, these CGMD models [25,26] mainly focus on the room temperature behavior of water or even sometimes the behavior below 0 ℃. While these models can capture freezing, ice formation and other properties of water [27], their applicability to high temperature problems is limited or seldom.

Simulating water heat transfer or evaporation using the MARTINI model in its original form will produce incorrect results; thus, it must be re-parameterized to temperatures near 100 °C, which is closer to the working temperatures of a semiconductor and the saturation temperature of water. The model is re-parameterized for higher temperatures using a hybrid ANN-GA-NM (Artificial Neural Network, Genetic Algorithm and Nelder Mead) approach and called it as MARTINI-E[28]. In a separate study the author of the current paper investigated[29] the popular potential functions like LJ, Morse, Mie and Stillinger-Weber (SW) for its suitability as a CGMD water model and found that Morse potential based models and machine learned SW (ML-mW) potential based models[27] shown good agreement with experimental values. ML-mW has an upper hand on Morse potential when it comes to accuracy and computational speed. However, the latter has good stability and less energy fluctuations with larger time steps of integration. Moreover, ML-mW is 1:1 mapping of water molecules to bead whereas Morse based, and MARTINI models are 4:1 mapping, which ultimately makes them much faster than ML-mW. Hence the studies in this paper are mainly based on the optimization of the water models based on Morse potential and its application on evaporation of water.

In the previous study, the author developed Morse-D model[29] which is parameterized for multi temperature values of water. Since the first paper was not published in a peer reviewed journal, the details of the same are reproduced in the current paper for completeness. Morse-D is re-parameterized to match the experimental density, enthalpy of vaporization and surface tension at room temperature (20°C aka 293 K as per NIST standards) and we call it as Morse-E. The optimization process starts with the Nelder-Mead method, followed by the Strength Pareto Evolutionary Algorithm. Our results of CGMD optimization studies shows that multi temperature optimization based models (Morse-D) yield more accurate results compared to single temperature models (Morse-E). In the second half of this paper, Morse-D is used for nanopore evaporation studies with varying pore diameter and temperature and important properties like heat flux and mass flow rates are estimated.

## Coarse Grained Models of Water

The most used coarse-grained model for water is MARTINI. Other interesting CGMD models of water include machine learned Stillinger-Weber (ML-mW) model[27] and extended MARTINI model (MARTINI-E[28]) (see supplementary material for details).

### Morse Potential

The Morse potential is a convenient inter-atomic interaction model for a diatomic molecule. It provides a better approximation for the vibrational structure of the molecule than the quantum harmonic oscillator [30] because it explicitly includes the effects of bond breaking, such as the existence of unbound states [31]. Though it was originally developed for diatomic molecules, the Morse potential can also be used to model other interactions such as the interaction between an atom and a surface, and recently for CGMD models. The Morse potential energy function takes the form as shown in Eqn. 1

$$\phi = D_0 \left[ e^{-2\alpha(r-r_0)} - 2e^{-\alpha(r-r_0)} \right] \tag{1}$$

$$E = \phi(r) - \phi(r_{cut}) - (r - r_{cut}) \frac{d\phi}{dr}\bigg|_{r=r_{cut}} \tag{2}$$

Here $r$ is the distance between the atoms, $r_0$ is the equilibrium distance, $D_0$ is the potential well depth, and $\alpha$ controls the spread of the potential function. For numerical simulations, often a shift and a linear term is added to the potential so that the potential energy and force both attain a value of zero at the cut-off radius (Eqn. 2). Generally, a cutoff radius ($r_{cut}$) is used to speed up the calculation using a computer, so that atom pairs with distances greater than the cutoff radius have an interaction energy of zero or negligible. The advantage of using Morse over the standard LJ potential is due to its versatile tunability using 3 parameters instead of 2. This will avoid sharp jumps in energy wells and can appropriately model the liquid properties near freezing point.

**Molecular System for CGMD simulation**

A lamellar system consisting of liquid film suspended between vapor layers is an ideal candidate to estimate the liquid density, vapor density, surface tension, enthalpy of liquid phase and enthalpy of vapor phase simultaneously. Such a system is constructed as shown in the Fig. 1a. To start with any optimization study, we need an equilibrated CGMD system. For that a 4.5 nm cubic periodic boundary system with MARTINI beads (Fig. 1a) equivalent to 1,000 $kg/m^3$ is equilibrated at 1 atm pressure and a temperature of 300 K using Nose-Hoover[32,33] thermostat and barostat. After equilibrating for 100,000 steps using a time integration step of 20 fs, the z-axis dimension is increased to 18 nm and sides are increased to 4.6 nm. This will form a lamellar system with periodic boundaries and consisting of 864 CGMD beads (equivalent to 3456 water molecules) as shown in the Fig. 1a and is then equilibrated for another 200,000 steps by removing the barostat and using only a Nose-Hoover[32,33] thermostat at 300 K. This equilibrated system is used for all further optimization trials of Morse-D and Morse-E models.

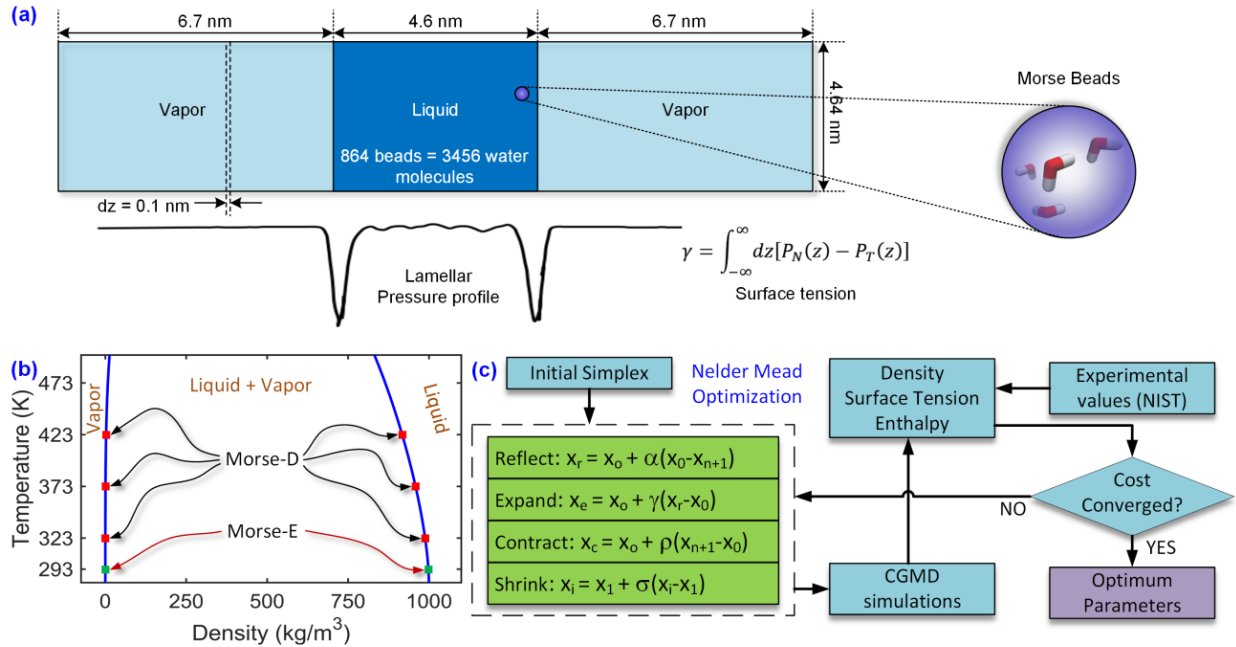

**Figure 1**: Lamellar CGMD system used to test the CGMD models. The system consists of equilibrated liquid film with MARTINI beads (initially) suspended between two vapor phases on either ends. Each bead in this system represent 4 water molecules. The expected sample pressure profile is shown with the surface tension ($\gamma$) calculation, with a slab thickness of 0.1 nm. b) Phase equilibrium points used to optimize the Morse-D (in red color) and Morse-E (in green color) models of water. c) Flow chart of the Nelder-Mead optimization method implementation.

To estimate the thermodynamic properties from a simulation, the lamellar system is divided into imaginary slabs of thickness $dz = 0.1\ nm$ as shown in the Fig. 1a. The number density, density, enthalpy, and surface tension are then estimated through the following equations.

The density profile is given by,

$$\rho(z) = m_{bead}\langle N(z)\rangle \qquad (3)$$

Here, the $m_{bead} = 72.0612\ \frac{g}{mol}$ is the CGMD bead mass and $N(z)$ is the number density in the $z^{th}$ slab. The enthalpy profile is given by,

$$H(z) = \langle KE(z)\rangle + \langle PE(z)\rangle + \langle P(z)V_{slab}\rangle \qquad (4)$$

Here, $H$ is the enthalpy, $KE$ is the kinetic energy, $PE$ is the potential energy, $P$ is the total pressure, and $V_{slab}$ is the volume of the slabs. The surface tension is given by,

$$\gamma = \sum_{z=0}^{z=L_z} dz [\langle P_N(z) \rangle - \langle P_T(z) \rangle] \tag{5}$$

Here, $L_z$ is the length of the lamellar system, $P_N$ and $P_T$ are the normal and tangential components of the pressure tensor at $z^{th}$ slab.

Figure 1b shows the two different approaches to optimize a force field. In the first approach, experimental values of water at 3 or more temperatures are compared with the simulation results and force field (Morse-D) is optimized. This is called multi-temperature approach for optimization. In the second approach, the simulation results are compared with the experimental values of water at a single temperature, and we call it as single temperature optimization.

**Stage-1 Optimization using Nelder-Mead Method**

One of the crucial steps in optimizing a potential or force field is the selection of initial bounds or the starting point for the parameters ($D_0$, $\alpha$, $r_0$) of Eqns. 1 and 2. For the development of Morse-E, the initial point was obtained from the multi temperature optimized model Morse-D[29] through a downhill simplex method called Nelder-Mead. This way, the simulation instabilities due to wide density fluctuations can be minimized. The density, surface tension and enthalpy of vaporization of water is estimated from the CGMD simulations and compared with the experimental results through a cost function (average percent error) as shown in the Eqn. 6. Here, $\delta$ is the average percent error; $\rho$, $H$, $\gamma$ are the density, enthalpy of vaporization and surface tension values from CGMD simulations. The quantities with a subscript of $EXP$ indicate the values from the experiments obtained from the NIST database [34].

$$\delta_{cost} = \frac{100}{3}\left[\left(\frac{\rho - \rho_{EXP}}{\rho_{EXP}}\right) + \left(\frac{H - H_{EXP}}{H_{EXP}}\right) + \left(\frac{\gamma - \gamma_{EXP}}{\gamma_{EXP}}\right)\right] \tag{6}$$

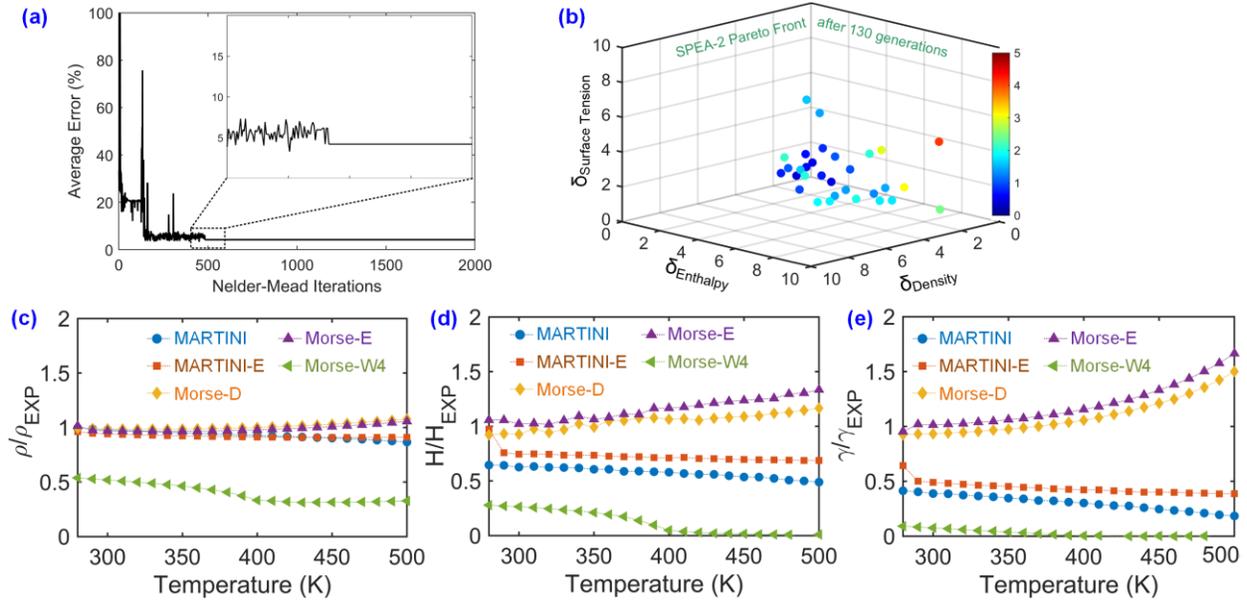

**Figure 2**: a) Average percent error of density, enthalpy of vaporization and surface tension for the Nelder-Mead method. b) Percent error for density, enthalpy of vaporization and surface tension is plotted as the pareto front of SPEA-2 after 130 generations. The color bar represents the average percent error of the three quantities. The ratio between the CGMD simulations and experimental values are shown for c) density, d) enthalpy of vaporization, and e) surface tension. The quantities are estimated for temperatures varying from 280 K to 500 K. The markers indicate MARTINI model (circle), MARTINI-E (square), Morse-D (diamond), Morse-E (triangle) and Morse-W4 (left triangle) models respectively.

The Nelder–Mead (NM) method is a commonly applied numerical method [35,36] to compute the minimum or maximum of an objective (cost) function in a multidimensional space. It is a direct search method and is often applied to nonlinear optimization problems for which derivatives may not be known. The method uses the concept of a simplex, which is a special polytope of $n + 1$ vertices in $n$ dimensions. The major steps of this algorithm are sorting the simplex based on the objective function values, reflection, expansion, contraction and shrinking. The simplex is iteratively progressed through multiple dimensions in direct search of the optimum parameters and terminated when the standard deviation of the simplex reached a threshold or upon reaching certain number of iterations. There exist many variations for the numerical implementation of the NM method, of which the current work uses the original method [35]. The algorithm of NM implementation is shown in Fig. 1c. Using this implementation, the Morse parameters are optimized to achieve the experimental values at 293 K.

The cost function (Eqn. 6) chosen for optimization tends to reduce the average error across three thermodynamic parameters. There is always a chance with algorithms like NM method of local convergence and thus we limit our computations for a certain threshold of the cost function. The results of the convergence are shown in the Fig. 2a. The inset figure shows the zoomed in error value region near 500$^{th}$ iteration. Apparently, the NM algorithm converged into a local minimum with an error value of 4.2% with $D_0 = 2.20926$, $\alpha = 0.59846$, $r_0 = 5.998474$.

To avoid this premature convergence metaheuristic techniques like genetic algorithm could be used. Since our objective is to minimize the error in 3 thermodynamic parameters ($\rho$, $H$, and $\gamma$), a multi objective pareto based method like Non-dominated Sorting Genetic Algorithm (NSGA) [37] or Strength Pareto Evolutionary Algorithm (SPEA) [38] will be appropriate.

**Stage-2 Optimization using Strength Pareto Evolutionary Algorithm**

Both NSGA and SPEA are best in solving multi objective optimization problems according to the literature. The latter is easier to implement due to its simplicity. This work used a python code implementation of the SPEA2 [38] an upgraded version of the original. Both these methods fall under the *A posteriori* methods of multi objective optimization techniques, which aims at producing all the Pareto optimal solutions. A detailed explanation of the steps and implementation details of this method is beyond the scope and focus of this work and hence readers are advised to refer the original paper [38].

The in-house developed python code is tested and validated for the standard functions (see supplementary material). for multi objective optimization This code is further extended to run the CGMD simulations and estimating the errors. The SPEA2 is started with a population size of 30, archive population of 20, mutation probability of 0.1, and the bounds of the variables are [1.5, 2.5] for $D_0$, [0.5, 1.0] for $\alpha$, and [5.0, 7.5] for $r_0$ respectively. All simulation runs are performed using a time step of integration of 20 $fs$, for 50,000 equilibration steps followed by 100,000 production steps using a Nose-Hoover (NVT) thermostat using LAMMPS software[39] unless otherwise specified. We call this model as Morse-E which represents the optimization at room temperature. The optimized pareto front after 130 generations is shown in the Fig. 2b. The axes correspond to the percent error for each thermodynamic variable and the color intensity represents the average percent error. After 130 generations of SPEA2, there was no significant improvement in error and found to 0.4 % with the parameters given in the Table 1. For comparison, the other Morse based model parameters are also listed in the Table 1. Note that the time step of integration and cutoff radius used for these models are different. For Morse-D and Morse-E, the cutoff radius is kept at 1.2 nm for faster performance and dt as 20 fs for low density and pressure fluctuations during the simulations.

**Table 1**: Optimized parameters of Morse based CGMD models

| CGMD Model | $D_0 (kcal/mol)$ | $\alpha (1/Å)$ | $r_0$ (Å) | $r_{cut}$ (nm) | $dt$ (fs) |
|---|---|---|---|---|---|
| Morse-D[29] | 1.686345 | 0.625349 | 5.8106 | 1.2 | 30 |
| Morse-W4 [40] | 0.81262 | 0.7 | 6.29 | 1.6 | 40 |
| Morse-E | 1.850222 | 0.683513 | 5.78284 | 1.2 | 20 |

The Morse-W4 model is aimed at representing the experimental density, enthalpy of vaporization, and surface tension at 298K. However, a larger cutoff radius of 1.6 nm was required for Morse-W4 for accurately represent the properties which makes it not so appealing. In the next set of studies, all the CGMD models are compared using same simulation conditions like cutoff radius of 1.2 nm, time step of 20 fs and using the same lamellar molecular system. The system is equilibrated for 50,000 steps, followed by 100,000 production steps and the temperature is controlled using a Nose-Hoover thermostat[32,33]. Five different CGMD models of water (MARTINI, MARTINI-E Morse-D, Morse-E and Morse-W4) are compared using the same simulation conditions. The temperature of the system is varied from 280 K to 500 K and properties are estimated. To account for the statistical fluctuations, 25 independent simulations are performed at every temperature with different starting conditions (random momentum of beads) and average quantities are estimated. These average quantities for density, enthalpy and surface tension are normalized with the corresponding experimental values and is plotted in the Fig. 2c, 2d and 2e respectively.

From the results shown in Fig. 2c, all models except Morse-W4 accurately represent density of water. The Fig. 2d shows the Morse-D perform better among all for enthalpy of vaporization. The surface tension results in Fig. 2e shows Morse-D as a better CGMD model among the selected four models. The superior performance of Morse potential based models for water with very short cutoff radius is promising to study thin film evaporation and wettability studies at nanoscale for temperatures around the boiling point of water. The poor accuracy of Morse-W4 is because original model used a cutoff radius of 1.6 nm, however we used 1.2 nm.

**Table 2**: Comparison of various models of water

|  | $\rho_{liquid}$ (298K) $g/cm^3$ | $dH_{vap}$ (298K) $kcal/mol$ | $\gamma_{LV}$ (300K) $mN/m$ | TMD (K) | $\rho_{liquid,MAX}$ (TMD) $g/cm^3$ |
|---|---|---|---|---|---|
| Exp [41] | 0.997 | 10.52 | 71.6 | 277 | 0.9999 |
| mW [42] | **0.997** | **10.65** | 66 | **250** | **1.003** |
| SPCE [16] | 0.999 | 10.76 | 61.3 | 241 | 1.012 |
| TIP3P [15] | 0.986 | 10.17 | 49.5 | 182 | 1.038 |
| Morse-W4 [40] | 0.998 | 9.17 | **71** | 50 | 0.941 |
| MARTINI [21] | 0.978 | 6.78 | 33 | 120 | 1.356 |
| Morse-D[29] | 0. 987 | 9.53 | 66.2 | 120 | 1.542 |
| Morse-E | 0. 978 | 10.29 | 73.3 | 120 | 1.478 |

More properties of these CGMD models are estimated and compared with other widely used water models including the atomic models like SPCE and TIP3P. The results are shown in the Table 2. The temperature at which the models show maximum density (TMD) is shown in the 4[th] column and the corresponding density is shown in the 5[th] column. Note that, SPCE and TIP3P models have explicit oxygen and hydrogens, mW considers one water molecule as a single bead and the rest of the models consider 4 water molecules as a single bead. The benchmarking studies to estimate the properties of water using various models at room temperature is performed and results are listed out in the table 2. The results show good performance of Morse based CGMD models and Stillinger-Weber potential based models in general. The simulation results from wider range of higher temperatures (Fig. 2c-e) concludes that the Morse-D and Morse-E performs superior in comparison with the rest of the CGMD models.

**Sensitivity of the time step of integration**

The time step of integration is an indicator of the CGMD models' sensitivity to represent the thermodynamic properties. To test this, the time step of CGMD models (MARTINI, Morse-D, Morse-E and Morse-W4) is varied from 2 fs to 62 fs and performed simulations for the lamellar system. This study is done for two temperatures (298 K and 373 K). The simulations start with a system equilibration for 50,000 steps using a time step of 20 fs and Nose Hoover thermostat. This step is followed by the production runs for 100,000 steps using corresponding time steps ($dt$) at micro–Canonical Ensemble (NVE). The resulting fluctuations in total energy and temperature is measured using the standard deviation of the data.

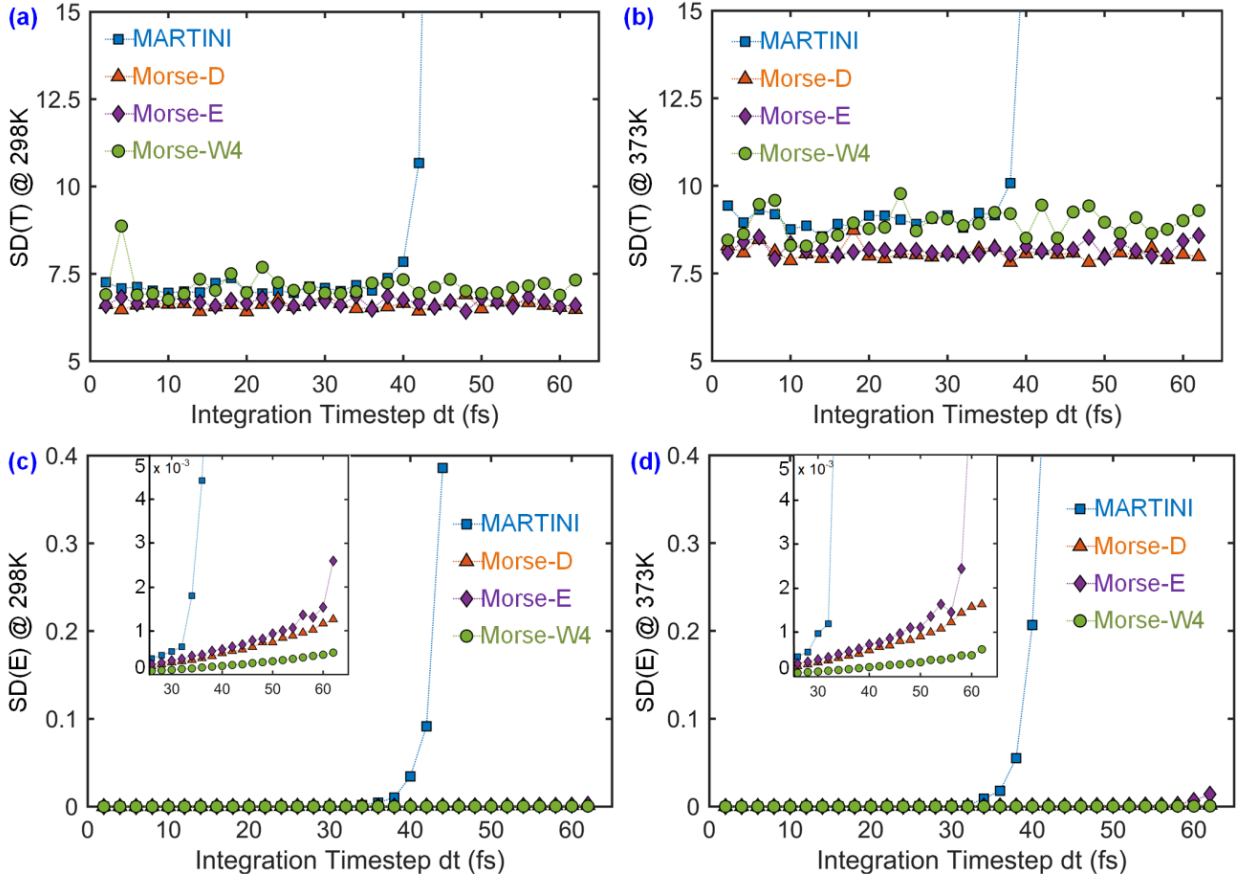

Figure 3: Standard deviation (SD) of temperature and total energy for various CGMD models is plotted against time step of integration. The results are taken from the micro canonical ensemble (NVE) simulation of the lamellar system equilibrated at 298 K and 373 K. Panels a) and b) is the result for SD of Temperature (SD(T)) and panels c) and d) represent SD of total energy (SD(E)). The inset shows the zoomed in view for the timesteps greater than 25 fs, which indicate a diverging trend of standard deviation of total energy of the system. The markers indicate MARTINI model (square), Morse-D (triangle), Morse-E (diamond) and Morse-W4 (circle) models respectively.

The standard deviation ($SD$) of the temperature and energy is estimated using the Eqn. 7 and Eqn. 8, where $T$ is the temperature of the system and $E$ is the total energy of the system with $N$ beads.

$$SD(T) = (\langle [T - \langle T \rangle]^2 \rangle / (N - 1))^{1/2} \tag{7}$$

$$SD(E) = (\langle [E - \langle E \rangle]^2 \rangle / (N - 1))^{1/2} \tag{8}$$

The results of the simulations and the fluctuations in temperatures are given in Fig. 3a (for 298 K), Fig. 3b (for 373 K) and the fluctuations in total energy are given in Fig. 3c (for 298 K) and Fig. 3d (for 373 K) respectively. The sensitivity study results show that the MARTINI model has the highest fluctuations in both energy and temperatures and is unreliable at high timesteps of integration. Moreover, the total energy diverges exponentially after 30 fs and the temperature diverges exponentially after 35 fs for MARTINI model. The fluctuations in total energy are small for Morse based models and hence the region from 25 fs to 62 fs is zoomed in (inset figures) for better clarity in Fig. 3c and Fig. 3d. Morse-D and Morse-E performs better than the rest for temperature fluctuations and Morse-W4 performs slightly better for total energy fluctuation. The maximum standard deviation of the total energy of the system using Morse based CGMD models is less than 0.2% of the equilibrium energy value ($D_0$) from Table 1 even at 50 fs or 60 fs. Hence a time step of integration between 30 fs to 50 fs can be used without any stability issues or divergence issues in energy and temperature of the system. Even though both Morse-D and Morse-E models performs best in the class, Morse-D shows a slightly better accuracy in representing thermodynamic properties.

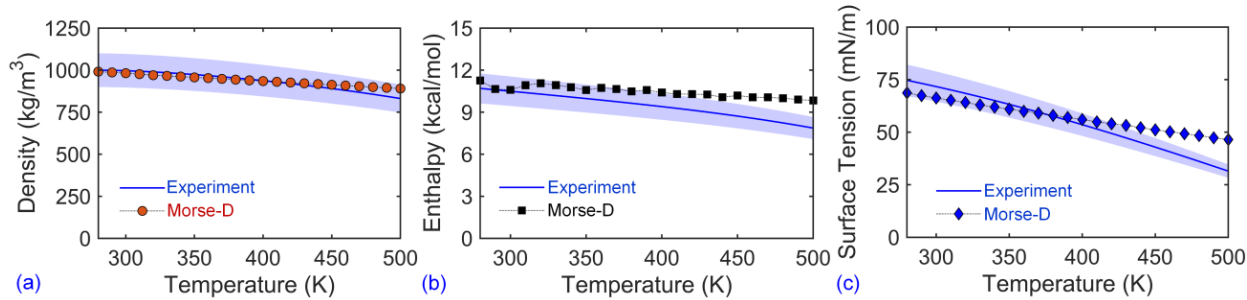

Figure 4: Comparison of Morse-D model's density (a), enthalpy of vaporization (b), and surface tension (c) with experimental data using a timestep of integration of 50 fs. The solid line indicates the experimental value as per NIST database and the blue shade indicates the 10% band. The standard deviation of simulation data is less than 2% in all cases and hence the error bars are not plotted.

The performance of the Morse-D model alone with respect to temperature is shown in the Fig. 4 for density, enthalpy of vaporization and surface tension. The Fig. 4a shows the density of the water is well represented by the model. The enthalpy of vaporization shifts away gradually from the 10% band of the experimental values (blue line and shade around it) as shown in the Fig. 4b. The surface tension values (Fig. 4c) indicate that the model's prediction deviates from experimental values only after $400\sim425\ K$. Hence this deviation is not an issue for systems with temperature reaching less than the boiling point of water $373\ K$. Due to the Morse-D model's slightly better accuracy at higher temperatures, it is used for all further studies including nanopore evaporation in this paper.

**Methods and tools used for force field optimization**

All CGMD simulations in this work are performed using the MPI version of LAMMPS software[43], version 15th April 2020. For visualization Visual Molecular Dynamics (VMD) software[44] is used. For CGMD simulations, a default time step of 20 fs is used unless otherwise specified. Python 3.9 programming language is used to create SPEA2 code, preprocessing and postprocessing tools for the CGMD simulations. C++ language-based code is developed and used for fast LAMMPS dump file reading and processing. A temperature of $293\ K$ is used for Morse-E parameterization since it is the room temperature according to NIST standards. Temperature of $298\ K$ is used to compare results with other water models. A temperature of $373\ K$ is used to check the models at boiling point of water.

## Nanopore Evaporation simulations

In this section we utilize the coarse grain water model Morse-D and a hydrophilic metallic nanopore to study the heat flux characteristics at nanopores. The cross-sectional schematic view of the system considered for studying heat flux is shown in the Fig. 5a. The red colored region is the nanopore with a diameter $d$. Modeling the complete solid region is a waste of computation and hence only the surface beads which interacts with Morse-D water beads (shown in red color) is modeled. Hence the pore is modeled with single layer simple cubic structure with a spacing of 0.5 nm between the beads in all directions. The water beads (in blue color) in the pore region are heated using a Langevin thermostat and is labeled in Fig. 5a as heating region. The heating of the water in nanopore is expected to evaporate upwards and the reservoir will supply water to support the wicking effect. Since the system is modeled with periodic boundary conditions, the upper and lower boundaries are connected. A cooling region of thickness 1.5 nm is considered at the top 15 nm from the nanopore opening. This will condense the vapor and supply it back to the reservoir. This will not violate any realistic physics near this region since the cutoff radius is 1.2 nm. The diameter of the pore is indicated as $d$ and other relevant dimensions of the system is labelled and shown in the left. For studying the pore evaporation, the diameter of the pore is changed from 2 nm to 5 nm and selected molecular models are shown in Fig. 5b-e. The heating of the pore region is expected to aid evaporation of the fluid as shown in Fig. 5f. In the Fig. 5f, the region shaded in green color is used to estimate heat flux and the region shaded in yellow is used to estimate mass flow rate.

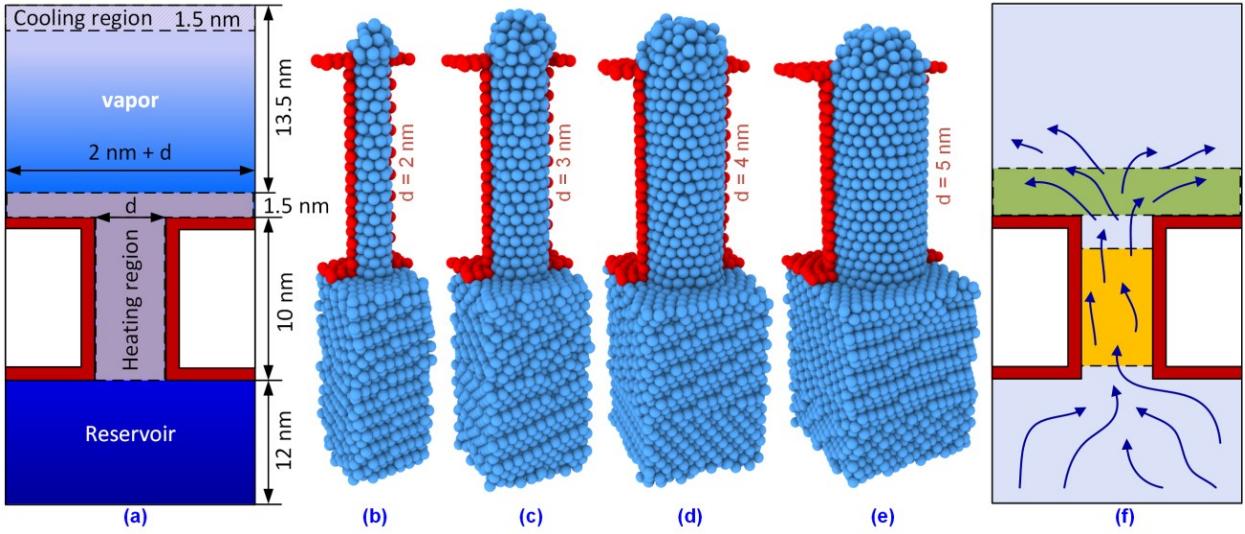

Figure 5: *In-Silico* system setup for the estimating the evaporative heat flux in nanopores. The conceptual cross sectional 2D layout of the periodic system with dimensions are shown on the left (a). The images of the molecular model with various nanopore diameters are shown in the middle (b-e). The nanopore (in red) is cut longitudinally to show the water beads (in blue) clearly. The heat flux estimation region (in green color) and the mass flow rate estimation region (in yellow color) is shown in the panel (f). The expected flow of the fluid in the system is shown in the schematic (right).

The water beads and solid beads interact through a Lennard-Jones potential (Eqn. 9) with $\sigma = 0.47\ nm$ and $\epsilon = 0.5\ kcal/mol$ corresponding to a hydrophilic surface[29].

$$E_{LJ} = 4\epsilon \left[ \left(\frac{\sigma}{r}\right)^{12} - \left(\frac{\sigma}{r}\right)^{6} \right] \tag{9}$$

Here, $\epsilon$ is the depth of the energy well, $\sigma$ is the equilibrium distance of beads and $r$ is the inter bead spacing.

The periodic system mentioned in Fig. 5 is modeled using a Python script with required dimensions and inter-bead spacings. Six nanopore systems with their diameters varying as 2 nm, 3 nm, 3.5 nm, 4 nm, 4.5 nm, and 5 nm is modeled and selected models are shown if Fig. 5b-e. Simulations are performed using the software LAMMPS [39]. The timestep of integration is $50\ fs$, cutoff radius of $1.2\ nm$, and the system was equilibrated for $1.25\ ns$ at $300\ K$ before any production runs were done. For the six systems with various pore diameters, the production runs are performed as explained next. The temperature of the heating region is connected to a Langevin thermostat which raises the temperature from $325\ K$ to $850\ K$ with a step increment of $50\ K$ every 2.5 ns and ran for $20\ ns$. Even though the Langevin thermostat is set at these ramped temperatures, the actual temperature of the water beads inside the pore will never achieve it due to the momentum loss in the presence of solid walls, and due to the enthalpy of evaporation which requires the system to be at saturation temperature while evaporating. These simulations are repeated 5 times with different starting configuration (momentum of beads) to perform statistical averaging of the results. Two of the main outcomes of these studies are the heat flux and the mass flow rate through the nanopore at regions shown in the Fig. 5f. The procedure for estimating these quantities is explained next.

**Estimation of Heat Flux from Molecular Dynamics simulations**

The estimation of macroscopic hydrodynamics properties from molecular simulations described by Irving and Kirkwood [45] date back to the 1950s. The original method required ensemble averaging which is computationally expensive. This is addressed through deriving expressions for instantaneous quantities by Evans and Morriss [46]. The final form of the instantaneous heat flux vector ($J_Q$) is derived as shown below.

$$J_Q(r,t) = \sum_i U_i v_i \delta(r - r_i) - \frac{1}{2} \sum_{i,j} r_{ij} F_{ij} \cdot [v_i + u(r_i) - u(r)] \times \int_0^1 d\lambda\ \delta(r - r_i - \lambda r_{ij}) \tag{10}$$

Here, $r$ is the position of interest in space, $r_i$ is the position of the $i^{th}$ atom, $U_i$ is the total energy of the atom, $v_i$ is the thermal velocity of the atom, $F_{ij}$ is the pair interaction force, $u$ is the streaming velocity, and $\lambda$ is the integration variable. For more details of derivation, formulation and terminologies, readers are referred to the original publication [46]. This expression has been modified to bring in computationally efficient virial stress calculation and simplified to accommodate the periodic boundary system [47] is estimated as per Eqn. 11.

$$\vec{J} = \frac{1}{V}\left[\sum_i \vec{v}_i\, e_i + \frac{1}{2}\sum_{i \neq j} \vec{r}_{ij}(\vec{v}_i \cdot \vec{F}_{ij})\right] \quad (11)$$

Here, $\vec{J}$ is the heat flux vector, $V$ is the volume of the control volume under consideration, $\vec{v}_i$ is the velocity and $e_i$ is the unit vector of the $i^{th}$ particle. The term $\vec{r}_{ij}$ is the particle separation and $\vec{F}_{ij}$ is the force between $i^{th}$ and $j^{th}$ particles. We use this equation to estimate the heat flux vector applied to the volume corresponding to the heat flux (as shown in Fig. 5f). From this vector, vertical component ($J_z$) is reported, since that is the component which resembles the removal of the heat from the nano porous surface.

**Estimation of Mass Flow Rate**

The mass flow rate ($\dot{m}$) at continuum level is defined as below equation.

$$\dot{m} = \rho A V \quad (12)$$

Here, $\rho$ is the density, $A$ is the projected area of cross section of the control volume, $V$ is the average velocity of the flow. Let the control volume is defined as shown in yellow color in the Fig. 5f, in between the elevation $h_1$ and $h_2$. If the diameter is $D$, then the volume is estimated as.

$$Vol = \pi D^2 (h_2 - h_1)\ [\text{Å}^3] \quad (13)$$

Let the ensemble average of the water beads in the control volume be $\langle N_w \rangle$. The mass of a single bead of Morse-D is $m_B = 72.06\, \frac{g}{mol}$ and Avogadro's number is represented as $N_A$. The projected area and mass of the water in the control volume is estimated as.

$$mass = \frac{\langle N_w \rangle m_B}{N_A}\ [g] \quad (14)$$

$$area = \pi D^2\ [\text{Å}^2] \quad (15)$$

$$\dot{m} = \frac{\langle N_w \rangle m_B}{N_A} \times \frac{1}{\pi D^2 (h_2 - h_1)} \times \pi D^2 \times V \quad (16)$$

The above expression for mass flow rate simplifies to the below equation.

$$\dot{m} = 119.66 \times \frac{\langle N_w \rangle V}{(h_2 - h_1)}\ [ng/s] \quad (17)$$

For all the quantities mentioned above, the units are consistent with the *real* units of LAMMPS software[39].

## Results and Discussion

Six molecular (CGMD) systems with nanopore diameter ranging from 2nm to 5nm is simulated and results are estimated as described in the previous section. The heat flux in the upper portion of the system is estimated based on the Eqn. 11 and plotted in the Fig. 6a. The markers indicate the ensemble averaged values of 5 independent simulations of the same system at regular time intervals of 2.5 ns. The trend shows an increasing heat flux near the saturation temperature of the water (100 ℃) and decreasing afterwards. The highest values of heat fluxes are observed for the 3 nm to 4.5 nm pore diameter cases. For the same set of studies, mass flow rate is estimated based on the Eqn. 17 and plotted in Fig. 6b. A similar trend as heat flux is seen in the case of mass flow too. However, the maximum value of mass flow rate is observed for the system with nanopore diameter 5 nm at temperatures closer to critical temperature of water.

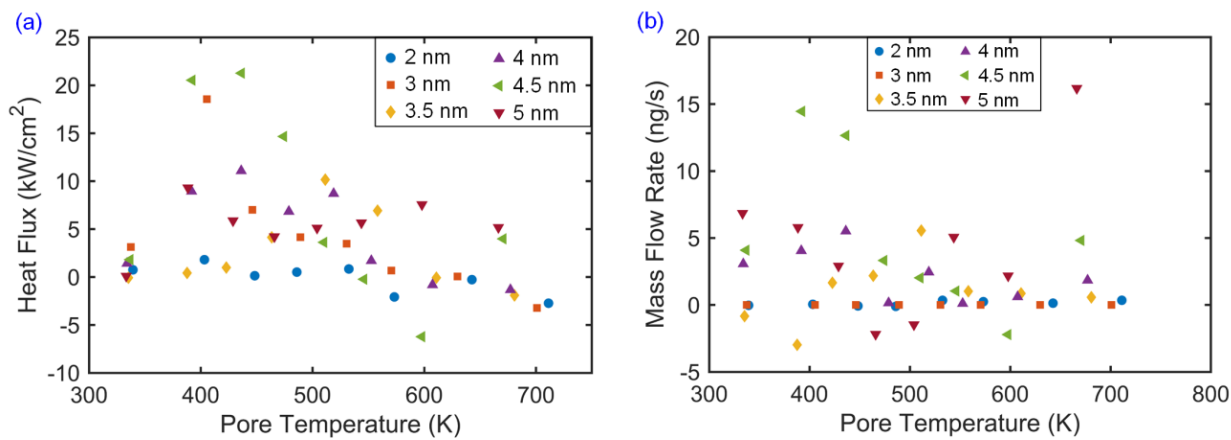

Figure 6: The estimated heat flux and mass flow rate for the nanopores is shown for diameters 2 nm (circle), 3 nm (square), 3.5 nm (diamond), 4 nm (triangle), 4.5 nm (left-tilted triangle) and 5 nm (inverted triangle). The variation of (a) heat flux and (b) mass flow rate is plotted against the temperature inside the nano pore.

Based on these obtained results, it is difficult to conclude a straightforward and simple trend. Hence, a pore burnout investigation is performed next. This involves identifying the cases and temperatures corresponding to partial or full pore burnout. A full pore burnout is defined as a situation in which the water evaporates from a nanopore leaving a hot substrate with no water in it to transfer heat. A partial burnout is when the liquid column in the nanopore is evaporated in the partial region (mostly upper region of the pore) which limits the continuous flow for evaporation. For our studies, molecular model trajectory is visually inspected using the VMD software[44] and inspected for partial or full burnout cases. Five sets of simulation trajectories (with different starting conditions) are inspected for each system (with a particular pore diameter) and for every pore temperature ($T_{pore}$). If the majority (3 out of 5) is showing a burnout then that is listed in the Fig. 7a. Note that in the present study no full pore burnouts are observed.

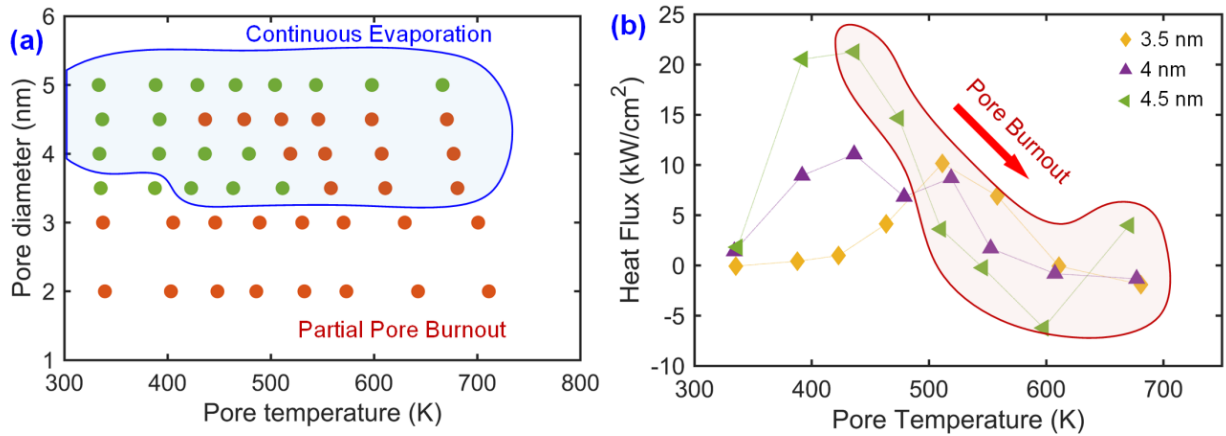

Figure 7: (a) Evaporation characteristics with respect to the diameter of the nanopore and the inner pore temperature is shown. The red color dots indicate partial pore burnout and green color dots indicate no burnouts in the system. The region marked using blue line facilitates continuous evaporation. (b) Heat Flux vs nanopore inner temperature is plotted for the selected 3 cases of nanopore diameters 3.5 nm, 4 nm, and 4.5 nm. The negative trend of heat flux is aligned with the partial pore burnout and is marked using red line in the figure.

The Fig. 7a shows the partial burnout conditions for various pore diameters and temperatures in red colored dots. In the same figure a region is marked as continuous evaporation region. A continuous evaporation region is defined using the simulation results when the nanopore is having a non-zero net positive mass flow rate over the given period. Despite having partial burnouts of the nanopore, for nanopore diameters 3.5 nm, 4 nm, and 4.5 nm a continuous mass flux is observed. To correlate this with the pore temperature and heat flux, these three cases are plotted in the Fig. 7b. The Fig. 7b shows a positive heat flux trend reaching to a maximum value and then receding to a lower value with a negative trend. This is in line with the pore partial burnout shown in Fig. 7a. We can conclude that the higher temperatures inside the nanopores are responsible for pore burnouts, at the same time heat flux is not increasing linearly with respect to the pore diameter. This contrasts with the intuition of heat flux will increase with increasing pore diameter. To address this confusion, further analysis is performed using the velocity profile inside the nanopore.

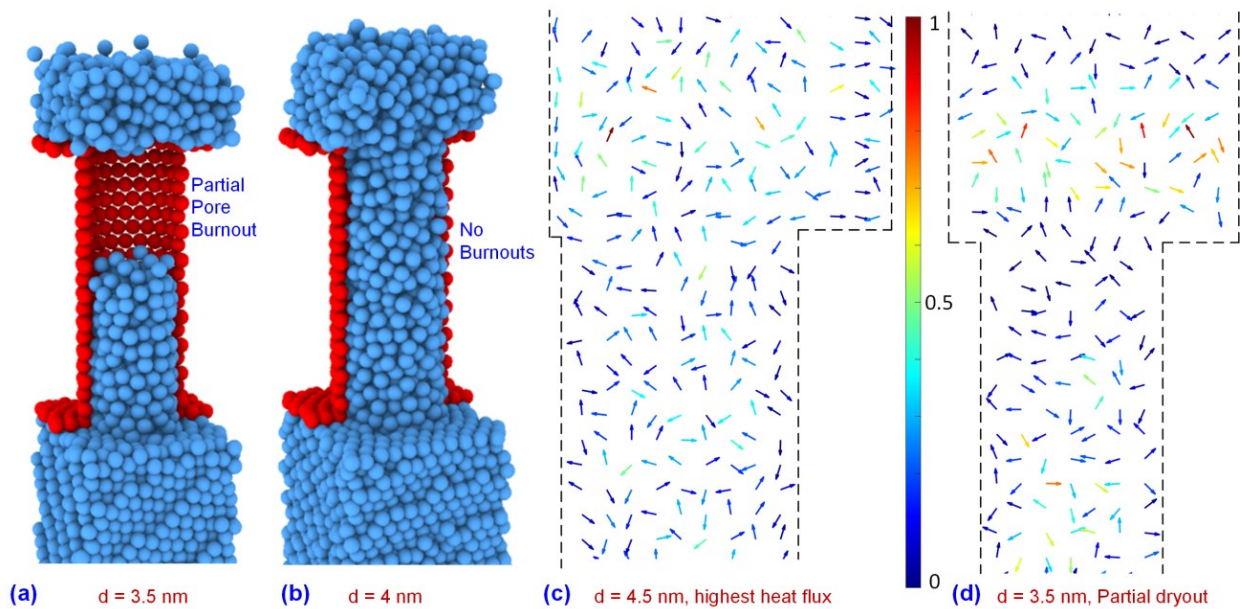

Figure 8: Snapshot of the coarse-grained MD system with (a) partial pore burnout at 558 K for pore diameter of 3.5 nm and a system without any burnouts is shown (b) for pore diameter of 4 nm at 478 K. Velocity profiles of the pore opening is shown in panels c and d. (c) velocity profile corresponding to the highest heat flux case (pore diameter of 4.5 nm, pore temperature of 436 K) and (d) velocity profile corresponding to the system with pore burnout condition whose pore diameter is 3.5 nm and pore temperature of 558 K. The velocities are normalized to unity and colorbar shows the value. Note that dark blue arrows have extremely small values and corresponds to burnout region.

A picture showing the difference between the partial pore burnout and no burnout situation is shown in Fig. 8a and Fig. 8b for pore diameters of 3.5 nm and 4 nm respectively. For the cases with continuous evaporation with partial pore burnout, the disconnected liquid column inside the pore will facilitate evaporating one bead at a time adding to the top liquid film. Snapshots of the molecular system at regular intervals for all the six cases with various pore diameters is shown in the supplementary document. Another interesting observation by comparing the Fig. 6a and 6b is that the heat flux and mass flow rate is not having the same trend even for the system with same nanopore diameter. To shed light on this, field-based velocity estimation in x-z plane is performed. The location of the molecule $(x_i, z_i)$ is converted to a field location $(x_g, z_g)$ by $(\left\lceil \frac{x_i}{dx} \right\rceil, \left\lceil \frac{z_i}{dz} \right\rceil)$. Here we choose $dx = dz = 2.5$Å, is the grid spacing of the field. At these grid points, the time averaging of the velocity is performed as $(\langle u \rangle, \langle v \rangle)$ for every 2.5 ns, where $u$ is the x-component and $v$ is the z-component of the velocities.

The averaged velocity profiles for a maximum heat flux case and a partial pore burnout case are plotted in Fig. 8c and Fig. 8d respectively. The values of the velocities are normalized to unity and the colorbar is shown in the figure. The dark color arrows represent grid points with minimum or negligible velocities. The location of the grid points (more or the right side compared to left side) gives an impression of a skewed geometry, but in fact it is due to the periodic boundary conditions and the ceil function used to calculate the grid points. The arrow with red color shows highest velocity and orange and yellowish colors shows some significant velocities of water molecules. For the case of partial pore burnout in Fig. 8d with a pore diameter of 3.5 nm, the velocity profile near the neck of the pore is consistent with the Fig. 8a. For the highest heat flux case, the velocity profile doesn't show a laminar flow like pattern as we see in continuum level simulations of fluid flow through pipes. Instead, the velocity profile shows a circulation pattern. The velocity in the center tends to go upward and recirculates around the walls and the same is observed in the top region of the nanopore. These results show that the flow patterns in nanopores are highly perturbed and non-laminar in nature with the presence of temperature gradient, instead of a pressure driven flow.

## Discussion and future directions

The inspiration to develop Morse potential based CGMD models was due to the undesirable artifacts produced by Lennard-Jones based models and promising the prior work by Molinero and team [42]. Following their route, force field parameters of CGMD water model based on Morse potential is fine-tuned using machine learning techniques. Two cases of optimization are considered, one at room temperature (Morse-E) and another at 3 temperatures (Morse-D) to match the enthalpy of evaporation, surface tension and density of water. For these optimization studies, self-diffusion coefficient was not considered. After the optimization studies, the results indicated that the Morse-D [29], a model that was done 2 years ago is better than Morse-E at representing the water properties at higher temperatures. Since the original manuscript of Morse-D was published in a preprint, the detail of model development is partially included in this paper and elaborated in the supplementary document.

While this work shows many interesting observations about the fluid flow and evaporation at nanoscale porous geometries, there is always room for improvement. The current study shows the critical diameter for the inception of evaporation in nanopores is in between $3\ nm$ and $4\ nm$. This study can be further extended by including the role of pressure characteristics and effect of hydrophobicity in evaporation dynamics. Another future direction can be investigation of role of long-range electrostatic forces which is missing in most of the CGMD models in evaporation simulations.

## Conclusions

In this study Morse potential based water models are investigated and optimized at a single temperature (Morse-E) and at multi temperatures (Morse-D). The optimization study results show that multi temperature based models like Morse-D performs better in simulating water evaporation at mesoscales. Using the Morse-D model, evaporation from hydrophilic nanopores with pore diameter varying from 2 nm to 5 nm is studied. Several interesting phenomena are observed, including the increasing-decreasing trend of heat flux with increasing pore temperature, continuous evaporation in nanopores despite of having partial burnout and disconnected liquid columns in pores, high heat flux and mass flow rate at lower temperatures for larger nanopores. The results also show that a pore diameter between 3 nm and 4 nm is the critical diameter to initiate continuous evaporation at nanopores. Both CGMD water models have very good numerical stability at higher time steps of integration and hence can perform longer simulations. This study gives insight into nanopore evaporation and serves as the basis for performing nanopore evaporation studies in complex geometries.

## Author Declaration

The author declares no competing financial interest.

## Supplementary Information

The supporting document consists of estimation of computational time required for evaporation of a small water droplet using MD simulations, various CGMD models of water, test functions for multi objective optimization, Nelder-Mead optimization of lamellar system, more details of Morse-D parametrization, and more graphical results of nanopore evaporation study.

Graphical TOC Entry

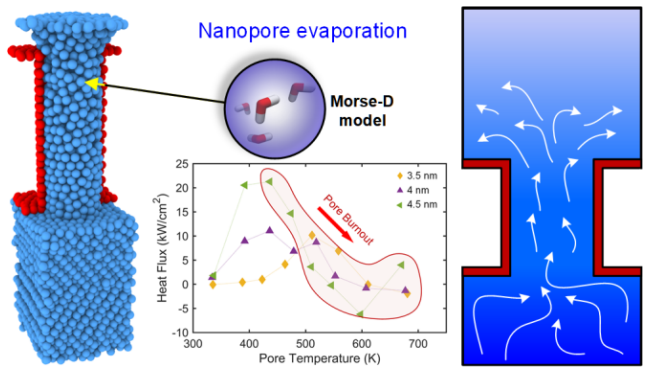